# Faster chiral versus collinear magnetic order recovery after optical excitation revealed by femtosecond XUV scattering

Nico Kerber[1,2]*, Dmitriy Ksenzov[3]*, Frank Freimuth[4], Flavio Capotondi[5], Emanuele Pedersoli[5], Ignacio Lopez-Quintas[5], Boris Seng[1,2,6], Joel Cramer[1,2], Kai Litzius[1,2], Daniel Lacour[6], Hartmut Zabel[1,2,7], Yuriy Mokrousov[1,2,4], Mathias Kläui[1,2]† and Christian Gutt[3]†

[1]Institut für Physik, Johannes Gutenberg-Universität Mainz, 55099, Mainz, Germany.
[2]Graduate School of Excellence Materials Science in Mainz, 55128 Mainz, Germany.
[3]Department Physik, Universität Siegen, Walter-Flex-Strasse 3, 57072, Siegen, Germany.
[4]Peter Grünberg Institute and Institute for Advanced Simulation, Forschungszentrum Jülich and JARA, 52425 Jülich, Germany.
[5]Elettra-Sincrotrone Trieste, 34149, Basovizza, Trieste, Italy.
[6]Institut Jean Lamour, UMR CNRS 7198, Université de Lorraine, 54506 Vandoeuvre-lès-Nancy, France.
[7]Department of Physics, Ruhr-University Bochum, 44780 Bochum, Germany.

*These authors contributed equally to this work
†Corresponding authors: christian.gutt@uni-siegen.de, klaeui@uni-mainz.de

## Abstract

While chiral spin structures stabilized by Dzyaloshinskii-Moriya interaction (DMI) are candidates as novel information carriers, their dynamics on the fs-ps timescale is little known. Since with the bulk Heisenberg exchange and the interfacial DMI two distinct exchange mechanisms are at play, the ultra-fast dynamics of the chiral order needs to be ascertained and compared to the dynamics of the conventional collinear order. Using an XUV free-electron laser we determine the fs-ps temporal evolution of the chiral order in domain walls in a magnetic thin film sample by an IR pump - X-ray magnetic scattering probe experiment. Upon demagnetisation we observe that the dichroic (CL-CR) signal connected with the chiral order correlator $m_z m_x$ in the domain walls recovers significantly faster than the (CL+CR) sum signal representing the average collinear domain magnetisation $m_z^2+m_x^2$. We explore possible explanations based on spin structure dynamics and reduced transversal magnetisation fluctuations inside the domain walls and find that the latter can explain the experimental data leading to different dynamics for collinear magnetic order and chiral magnetic order.

**Introduction**

In the field of magnetism and spintronics chiral magnetic structures, such as spin spirals, domain walls and skyrmions [1-6], are intensively investigated due to their fascinating properties such as potentially enhanced stability and efficient spin-orbit torque driven dynamics [7-10]. It has been shown that these structures are stabilized by the Dzyaloshinskii-Moriya interaction (DMI) [11-12] that favours a chiral winding of the magnetisation. This antisymmetric indirect exchange interaction requires materials with large spin orbit coupling as well as a broken inversion symmetry, present in special bulk systems such as B20 compounds where skyrmions have been first discovered experimentally [1-2] or in interfacial systems such as heavy metal/ferromagnet multilayer stacks [3-6, 8, 10]. The domain wall type (Néel or Bloch) and chirality of the spin textures can be accessed in real space by imaging techniques [4, 13-15] or in reciprocal space by (resonant) magnetic X-ray scattering [16-18]. The chiral wall spin structure is of key importance as it governs the dynamical properties of domain walls and skyrmions [7-10, 19]. While the investigation of static structures and slow (ns) dynamics of chiral magnetic structures has been intensified recently, experimental studies addressing the ultimate fs-ps dynamics of the chirality have been elusive so far. Ultrafast pump-probe experiments have concentrated on the collinear order in magnetic systems [20-28] with also some studies of non-collinear order in antiferromagnets [29-30]. In particular, as the ferromagnetic alignment minimizes the Heisenberg exchange energy, while the chiral order is resulting from the DMI, the ultrafast dynamics of both orders needs to be probed individually. And since the characteristic time scale for the onset of the chiral magnetic order and its ultrafast dynamics are unexplored up to now, we need to ascertain both, as they hold fundamental insights into the underlying physical mechanisms and allow us to gauge the ultimate speed for the manipulation of chiral magnetism, e.g. for ultrafast writing of chiral spin textures.

In this work we employ circularly polarized light pulses from an XUV free-electron laser and investigate time-resolved the evolution of the chirality of domain walls in magnetic thin film samples by an IR pump - X-ray magnetic scattering (XRMS) probe experiment.

Using samples with interfacial DMI and perpendicular magnetic anisotropy exhibiting labyrinth-like domain patterns, we measure in the same experiment both the sum signal corresponding to the ferromagnetic order in the domains and the difference signal corresponding to the average chiral order in the domain walls. We find an ultrafast intensity decrease of both signals in the sub-ps regime with similar time constants. However, a significantly faster recovery of the chiral signal in the sub-ns timescale is observed. We subsequently investigate the origin of the faster recovery of the chiral signal by performing numerical simulations of the scattering signal, which reproduce the experimental findings.

## Results

**Magnetic small angle X-ray scattering**

The experimental setup for the scattering experiment is shown schematically in Fig. 1a. Circularly polarized extreme ultraviolet (XUV) radiation of 60 fs pulse duration was tuned to a wavelength of 23.0 nm corresponding to the Fe $M_{2,3}$ dichroic transition, which exhibits magnetic scattering contrast due to the X-ray magnetic circular dichroism (XMCD) effect [27-28, 31-32]. The limited transmission in the XUV regime required us to perform the experiment in reflection geometry with an incident angle of $\Theta = 44°$ [28] yielding an effective XUV penetration depth of 8 nm. Due to an isotropic disordered domain structure we observe the typical ring-like diffraction feature well known from magnetic transmission SAXS experiments representing a broad distribution of Fourier components of the magnetic domain pattern [27-28]. The sample is pumped with a 60 fs IR laser pulse of 780 nm wavelength impinging on the sample with a small 2 degree offset with respect to the XUV beam (further setup information can be found in the Methods section).

The reflection SAXS pattern was detected on a 2D CCD-detector and investigated as a function of the pump-probe delay over a time span of 100 ps. For each time delay and helicity, 7000 scattering patterns have been measured, normalized to the incoming flux and averaged. The area around the beamstop and charge scattering streaks have been masked. The background from the charge scattering due to the reflection geometry has been subtracted for each pattern leading to the corrected patterns used in the further analysis.

The investigated [Ta(5.3 nm)/$Co_{20}Fe_{60}B_{20}$(0.93 nm)/Ta(0.08 nm)/MgO(2.0 nm)]$_{x20}$/Ta(1.6 nm) multilayer stack was produced by dc magnetron sputtering and exhibits perpendicular magnetic anisotropy (PMA). The material was imaged via magnetic force microscopy (MFM) as shown in Fig. 1b and exhibits, after out-of-plane demagnetisation, labyrinth-like magnetic domains at zero field. The domain pattern extended through all ferromagnetic layers due to stray field coupling exhibits a domain periodicity of (455±30) nm. In reciprocal space the MFM images yields a first order peak at q = (13.8±0.9) $\mu m^{-1}$. Due to partial alignment of the magnetic stripes, higher order peaks are visible as well [33]. Further material analysis with a superconducting quantum interference device (SQUID) yields a saturation magnetisation of $M_s$ = (844±28) kA/m and an effective perpendicular anisotropy $K_{eff}$ = (133±17) kJ/m$^3$ of the sample at room temperature (for more sample details see Methods section). The Ta(5.3 nm)/$Co_{20}Fe_{60}B_{20}$ interface leads to a positive DMI, with reported values for such stacks ranging from 0.06 to 0.30 mJ/m$^2$ [34-36], which favours right-handed Néel-type domain walls (see Fig. 1c). The exact domain wall arrangement is finally determined by an interplay between interfacial DMI and dipolar interactions, which can lead to complex domain wall arrangements as shown recently [37]. We also note that considering the small penetration depth of the XUV radiation, we are sensitive to the magnetisation of the topmost $Co_{20}Fe_{60}B_{20}$ layer only.

X-ray scattering from such magnetic structures leads to the ring shaped SAXS patterns shown in Fig. 1d. We facilitate the discussion by referring to expressions of scattering signals from a one-dimensional magnetic domain arrangement with perpendicular magnetization in the z-direction, lattice spacing d and Neel-type domain walls with a rotation into the x-direction. For this case the sum signal from intensities measured with circular left (CL) and right (CR) polarized light is given by (details see methods)

$$I^{CL+CR}(Q) = 2\left|\sum_n e^{iQdn}\right|^2 [|m_z(Q)|^2 + |m_x(Q)|^2] \quad (1)$$

with $m_{z,x}(Q)$ denoting the Fourier transform of the magnetization profiles. The strength of the $I^{CL+CR}$ signal is thus predominantly connected to the average domain magnetisation magnitude (I $\propto$ M$^2$) [27, 38]. Fig. 1d displays the corresponding sum image of the CL and CR signals showing a ring-like diffraction pattern at a q-value of (14.3±0.1) $\mu m^{-1}$ that we assign to the 1$^{st}$ order scattering peak of the magnetic stripe domains.

The sensitivity to the chirality of domain walls is obtained via the dichroic (CL-CR) scattering signal [17]

$$I^{CL-CR}(Q) = 4\left|\sum_n e^{iQdn}\right|^2 [\Im(m_x(Q))\Re(m_z(Q))] \quad (2)$$

the strength of which is given by the correlator $\Im(m_x(Q))\Re(m_z(Q))$ between z- and x-components of the magnetization. We also note that, in contrast to the CL+CR signal, phase information is preserved in the dichroic signal allowing to determine the domain wall chirality (left/right-handed) and character (Bloch (helical), Néel (cycloidal)) [17-18].

The measured dichroic signal (CL-CR) is shown in Fig. 1e and exhibits a pronounced angular asymmetry with an amplitude of 14 % of the sum signal. Fig. 1f shows the orthoradial profile of the dichroic signal. We used a fitting model of $A*\cos(\Psi-\varphi)$, where $A$ is the amplitude, $\Psi$ the azimuthal angle and the phase $\varphi$ determines the domain wall angle and with that the domain wall chirality and character [17]. Fitting the azimuthal profile leads to a value of $\varphi = 90°$ which indeed confirms the predominance of fully right-handed Néel-type domain walls [17-18] in the uppermost magnetic layer as expected by micromagnetic simulations of the full material stack in the Supplementary S3. This provides us a tool to individually probe the time resolved dynamics of the chiral magnetic order in the domain walls.

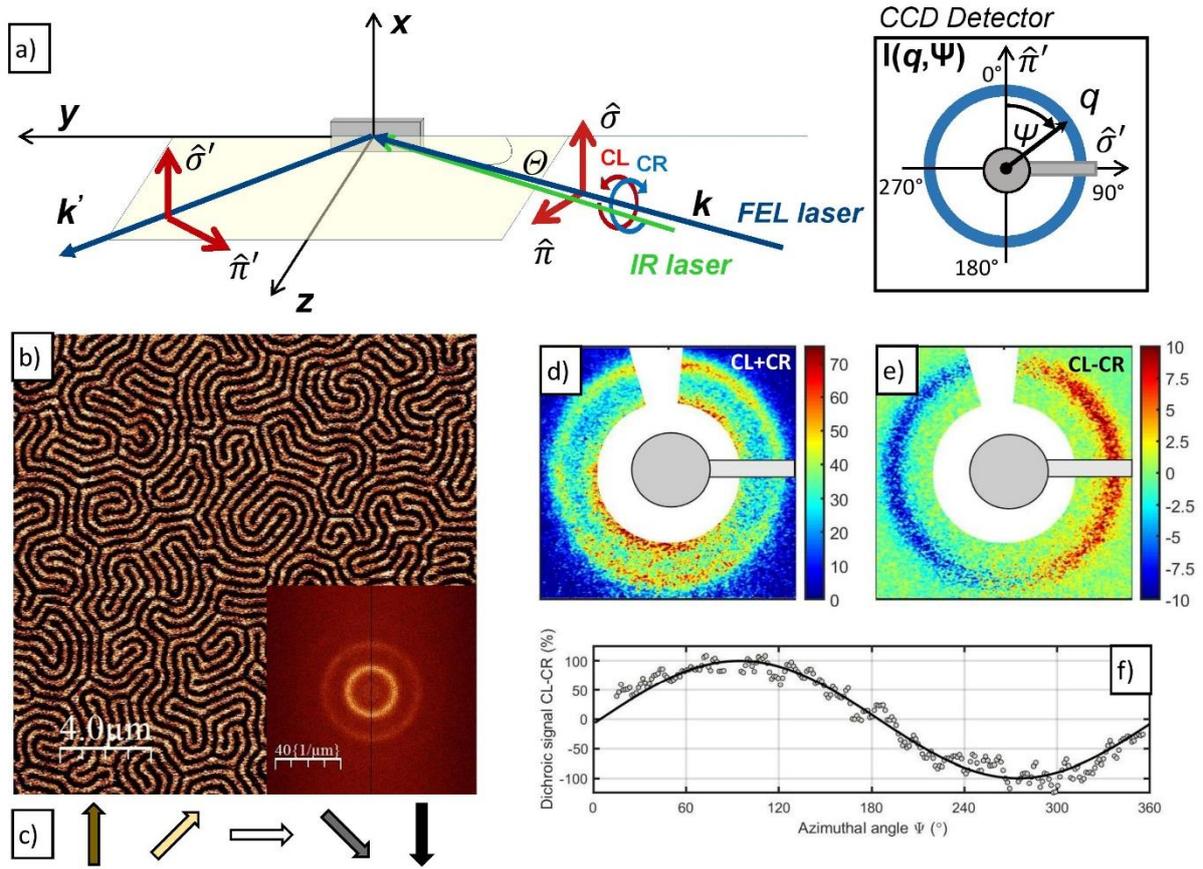

**Figure 1: Experimental setup and diffraction images.** (**a**) Measurement geometry: A magnetic thin film sample is pumped by an optical infrared laser pulse and probed by a circularly polarized X-ray FEL pulse with an incident angle of $\Theta = 44°$ (wavevector k, s-polarisation($\sigma$) and p-polarisation($\pi$)) that scatters on the sample. Afterwards an IR-protected charge-coupled device (CCD) detector records the magnetic SAXS pattern. (**b**) MFM image of a typical labyrinth domain pattern of the [Ta(5.3 nm)/$Co_{20}Fe_{60}B_{20}$(0.93 nm)/Ta(0.08 nm)/MgO(2.0 nm)]$_{x20}$/Ta(1.6 nm) sample. The inset shows the FFT with the first order peak corresponding to isotropic distributed labyrinthine stripes with a domain periodicity of (455+/-30) nm. These magnetic structures lead to the SAXS signals for left-hand (CL) and right-hand (CR) circular polarized incident x-rays. (**d**) The resulting sum = CL+CR (Eq. (1)) of the diffraction pattern confirms that the diffraction corresponds to the magnetic domains observed by MFM. (**e**) The dichroic scattering signal = CL-CR (Eq. (2)) and its azimuthal dependence (**f**) confirms the presence of (**c**) right-handed chiral Néel (cycloidal) domain walls.

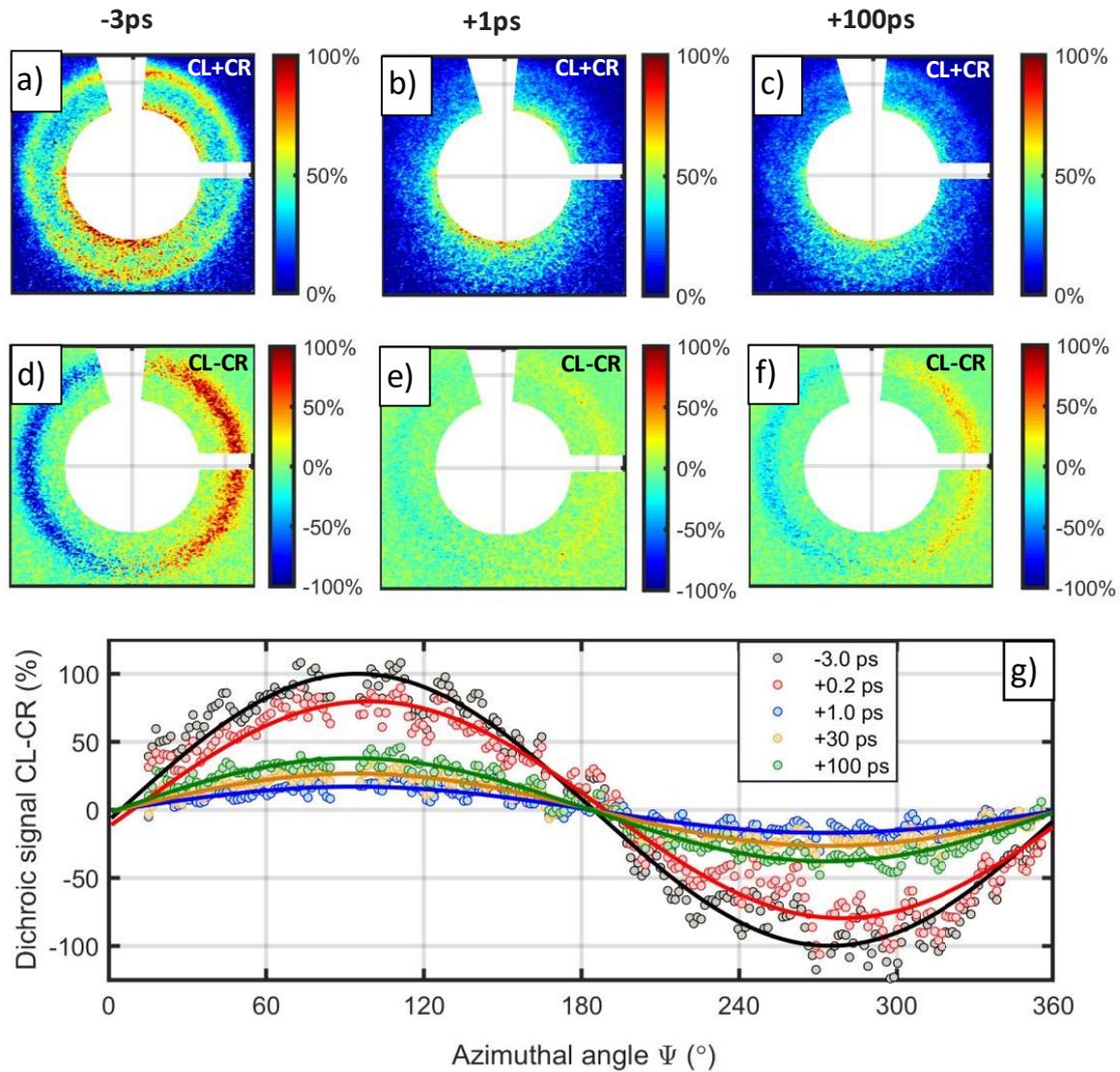

**Figure 2: Time dependence of diffraction images.** Diffraction patterns for the sum CL+CR (*top*) and the difference CL-CR (*bottom*) images for the initial (**a, d**), +1.0 ps (**b, e**) and +100 ps (**c,f**) states. (**g**) The experimental (*dots*) and fitted (*lines*) orthoradial dichroic profiles for different pump-probe delay times showing a maximum at 90° and a minimum at 270° corresponding to a right-handed cycloidal winding of the domain walls. While the amplitude of the dichroic signal drops within a ps upon pumping and recovers afterwards, no significant angle shift in the fitted phase φ can be detected. The beamstop is located at an azimuthal angle of 90°.

**Ultrafast time resolved pump-probe experiments**

In a next step we investigated the time evolution of the scattering signals upon IR laser excitation. Fig. 2(a-f) shows the diffraction patterns of the sum and the dichroic signals at different pump-probe delays. At +1.0 ps (Fig. 2b & 2e)) the intensity in both signals has dropped down significantly in comparison to the initial states (Fig. 2a & 2d) with a visible partial recovery of both signals after +100 ps (Fig. 2c & 2f).

By azimuthal integration of the dichroic signal we can identify that within our experimental accuracy no significant shift in the phase φ of the orthoradial profile (Fig. 2g) has been observed upon pumping. Thereby we can exclude any significant ultrafast change of the domain wall angle. This means that the DMI is even at the ultrafast timescale contributing to the wall chirality and therewith the stability of chiral spin structures is retained.

The key step is the comparison of the ferromagnetic order dynamics and the chiral order dynamics. To ascertain both, we calculated the average intensity for the sum and the difference signal as a function of the wavevector $Q_r$ for different time delays (Fig. 3a & 3b). The single $Q_r$ points shown represent averages of > 3000 pixels. Averaging these signals in addition along $Q_r$ (radial average) represents averages over $7.1 \times 10^5$ pixels containing in total between $1 \times 10^5$ and $5 \times 10^5$ photons. With this the numerical integration of the radial profiles for each time delay leads to the data shown in Fig. 4a which demonstrates the evolution of the total intensity of the sum and the difference signal as a function of delay time normalized to the unpumped total intensity. The data shown in Fig. 4b is the result of a second experimental run and represents the average of three scans shown in the Supplementary S9. We discuss the details of errorbar determination in the Supplementary S8 and note here that the maximum error is on the order of 1% for the data in Fig. 4a and 2 % for the data shown in Fig. 4b.

We note that the time constant of the ultrafast decay of the magnetization is similar for both signals (details in Supplementary S7). However, we observe a stronger demagnetization of the CL-CR signal which drops to (15±1)% at t=1 ps while the CL+CR signal drops to a value of (20±1)% only (Fig. 4a), similar values can be seen in Fig. 4b. The obtained data in Fig. 4a demonstrates that the CL+CR signal recovers by an amount of (8±2)% from 1 ps to 100 ps while the CL-CR signal recovers in the same time interval by (26±2)%. Similar values for the second experimental run can be seen in Fig. 4b. We used the model reported in Eq. (3) of the Methods section to fit the experimental data [27]. The model includes the time constant $\tau_d$ of the demagnetisation process of the ferromagnetic collinear order respectively the dechiralisation process (vanishing of the chiral order in the domain walls), and the time constants $\tau_{r1}$ and $\tau_{r2}$ of the fast and slow recovery processes. The initial ultrafast demagnetisation time constant $\tau_d = 0.39\pm0.10$ ps is in agreement with the typical demagnetisation times of ferromagnetic 3d-transition metals [21-23, 27-28]. The main results of these experiments are firstly a dechiralisation process ($\tau_d = 0.31\pm0.10$ ps) that occurs on similar timescales as the demagnetisation of the collinear domains, and secondly we observe a clear difference in the slower recovery process of the two signals. The time constant $\tau_{r2} = 312\pm18$ ps of the difference signal is significantly smaller than the sum signal ($\tau_{r2} > 900$ ps) demonstrating a faster recovery of the dichroic signal that reflects the chiral magnetic order in the domain walls.

To understand the origin of this difference in the recovery, we next investigate the time evolution of the 2nd moments of the radial distributions as displayed in Fig. 4 c & 4d. The 2nd moments (calculated using Eq. (4) in the Methods section) entail additional information about possible ultrafast changes of the radial distribution widths upon IR excitation. The 2nd moment of the sum signal increases significantly upon pumping within 1 ps, indicating a wider radial distribution and therefore we can identify stronger fluctuations leading to a reduced correlation length of the domain-domain correlation function. After 100 ps it has just recovered partially to its unpumped value. In contrast, the 2nd moment of the difference signal decreases slightly within 1 ps to about 90 % after optical excitation but recovers afterwards within a few ps to values around the unpumped value. These differences between the 2$^{nd}$ moments of CL+CR and CL−CR may be understood from the 2$^{nd}$ moments $\Delta I^{CL}$ of CL and $\Delta I^{CR}$ of CR. Defining $\Delta I_1 = [\Delta I^{CL} + \Delta I^{CR}] / 2$ and $\Delta I_2 = [\Delta I^{CL} − \Delta I^{CR}] / 2$ the 2$^{nd}$ moment of the sum signal is $\Delta I^{CL+CR} = \Delta I_1 + \Delta I_2 \, I^{CL-CR} / I^{CL+CR}$, while the 2$^{nd}$ moment of the difference signal is $\Delta I^{CL-CR} = \Delta I_1 + \Delta I_2 \, I^{CL+CR} / I^{CL-CR}$. In our case $I^{CL-CR} / I^{CL+CR}$ is of the order of 10%. Therefore, $\Delta I^{CL+CR}$ is rather insensitive to $\Delta I_2$. However, $\Delta I^{CL-CR}$ is very sensitive to $\Delta I_2$, because it is amplified by the inverse magnetic ratio $I^{CL+CR} / I^{CL-CR}$. When $\Delta I_2$ becomes negative, due to a different broadening of the CL and CR signals, $\Delta I^{CL-CR}$ may therefore decrease. Additionally, we point out that a simple reduction of the magnetization by a single demagnetization factor everywhere in space reduces both $I^{CL+CR}(Q)$ and $I^{CL-CR}(Q)$ by the same factor.

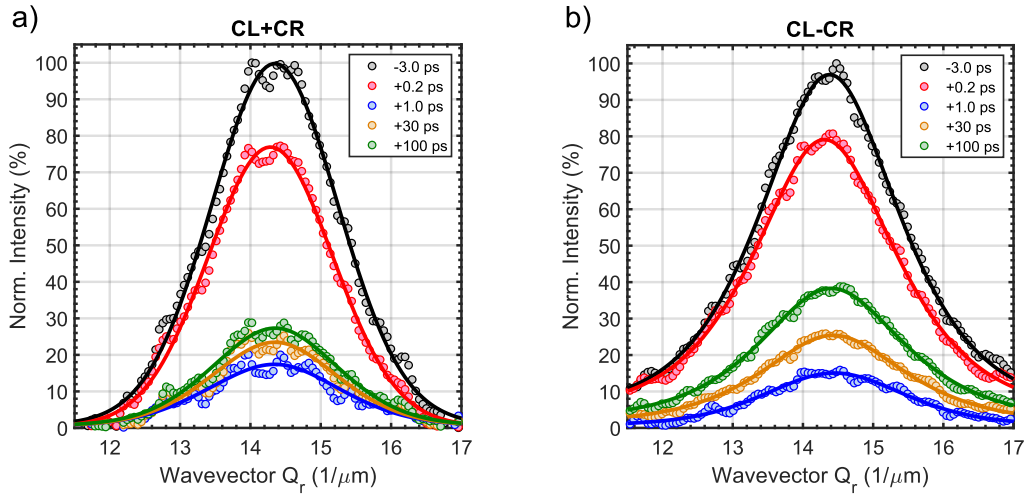

**Figure 3: Time dependence of radial scattering intensities.** The radial profiles of the sum CL+CR (**a**) and difference CL-CR (**b**) signals for selected delay times. Error bars from counting statistic are smaller than the symbol sizes. The visible fluctuations in the profiles are due to speckles originating from the coherent illumination.

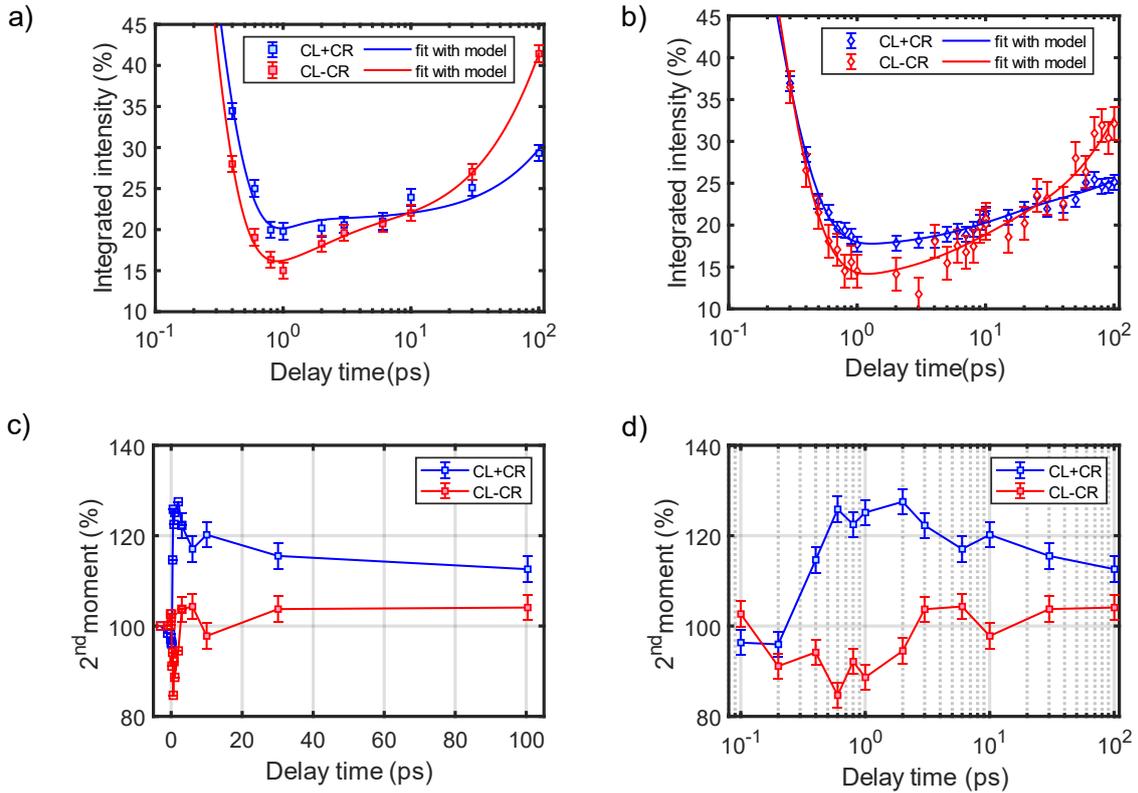

**Figure 4: Time dependence of total scattering intensities and second moments.** Time evolution of the (**a,b**) radially integrated intensity from two different experimental runs and (**c,d**) the 2$^{nd}$ moments of the sum signal (ferromagnetic order) and the difference signal (chiral magnetic order) from the scan shown in (**a**). The solid lines in (**c,d**) provide a guide to the eye.

**Discussion**

The experimental results show an ultrafast decrease within the first ps of the total intensity of the sum signal that reflects the conventional collinear magnetic order inside the domains, as well as the dichroic signal that corresponds to the chiral order of the domain walls. Both signals decay on similar timescales, indicating that for the ultrafast initial demagnetisation process the order of the magnetisation itself does not play a major role in our sample system and thus spin-flip probabilities are similar for both orders.

In order to explain the key finding of the experimentally observed faster recovery of the chiral signal after laser excitation we can envisage two different mechanisms: (1) a change in the size ratio between domain walls and domains caused by an increase of the domain wall width during the whole investigated time frame or (2) a faster recovery dynamics of the chiral order within the domain walls compared to the ferromagnetic order in the domains leading to a faster build-up of the chiral magnetisation.

The increase of the domain-wall width as required for mechanism (1) can result from the temperature dependence of the micromagnetic parameters, more precisely the effective anisotropy and the exchange constant, as shown in the Supplementary S4. Since the average domain magnetisation is directly proportional to the square root of the total intensity of the sum signal one can estimate the temperature evolution of the spin system (assuming the 3-temperature model [21]) during the pump-probe experiment using the fitted temperature dependence of the saturation magnetisation of the sample (see Supplementary S2). This leads to an increased spin system temperature of (493±22) K at a delay time of 1 ps, where the sample is significantly demagnetized, and an already decreased temperature of (469±13) K at 100 ps. Such a sudden change of the micromagnetic parameters is expected to excite a breathing mode of the domain walls. The breathing mode corresponds to a damped ps-scale oscillation of the width of the domain wall around the new equilibrium wall width as determined by the temperature-scaled micromagnetic parameters. We estimate the equilibrium wall width at the increased spin system temperature to be 11 nm (see Supplementary S4) in comparison to the equilibrium width of 8nm at room temperature determined in Supplementary S3. Therefore, we estimate that for our system the wall width oscillates between values of 8 nm and 14 nm due to the sudden change of the micromagnetic parameters. Domain-wall breathing frequencies lie in the range of several GHz. Assuming a breathing frequency of 5 GHz, one-half period of the breathing mode amounts to 100 ps, in which the wall expands by 6 nm from 8 nm to 14 nm.

Mechanism (2) is based on a faster recovery of the chiral order parameter of the system in comparison to the collinear order parameter [39-41]. The chiral order parameter is defined here as the in-plane component of the vector $\langle S_i \times S_{i+1} \rangle / |\sin \theta_0|$ with $\theta_0$ denoting the average angle between the neighbouring spins $S_i$ and $S_{i+1}$ in the wall [39-41]. The chiral order parameter defined in this way is a pseudo-scalar variable analogous to an Ising spin, which takes values of +1 for right-rotating walls and -1 for left-rotating walls, before the pump pulse arrives.

Between 1 ps and 10 ps the electronic part of the system recovers its ground-state properties to a large extent, while the spin system is still excited with a relatively high spin temperature of about 490 K, which slowly decays on the scale of 10 - 1000 ps towards room temperature. It is in this time window that the recovery rate of the chiral order parameter is significantly higher than that of the magnetization. This is based on different temperature dependences of the chiral and scalar order parameters as observed in various spin systems, with the chiral order parameter restoring to its ground state value faster as the temperature is lowered [41]. The faster recovery behaviour can be traced back to the properties of the chiral order parameter as an essentially Ising type variable [39-40, 42], which is also the basis of exotic spin states such as chiral spin liquids [39-40, 43].

In order to check the appropriateness of these two mechanisms for our case, we performed numerical simulations of the scattering signal and evaluated - similar to the experiment - the total intensity and the $2^{nd}$ moments of the sum and difference signal under the influence of these two mechanisms. We start from an initial labyrinth stripe state shown in Fig. 1b with a domain periodicity of 450 nm and a domain wall width of 8 nm as estimated in the micromagnetic simulations (Supplementary S3). For mechanism (1) the domain wall width was altered, while keeping the domain periodicity fixed. Mechanism (2) was implemented by introducing smaller amplitudes of the (transversal) magnetisation fluctuations to the domain walls compared to the domains emulating the faster build-up of the chiral order parameter.

In Fig. 5a we show the simulated total scattering intensities of CL+CR and CL-CR as a function of the domain wall width. For an estimated maximum increase of the wall width towards 14nm the signal strength of CL-CR increases to 174%, while the CL+CR increases slightly to 100.8%. The ratio of both signals thus increases by a factor of 1.7, which is considerably larger than the experimental value of ≈1.3 as measured at 100 ps (Fig. 4a,b).

In Fig. 5b we show the dependence of the simulated total scattering intensities of CL+CR and CL-CR on the fluctuation tilt angle, which characterizes the strength of the transverse spin fluctuations. The dashed and dotted lines in Fig. 5b correspond to a homogeneous distribution of fluctuations throughout the system, while the solid lines represent the case of reduced fluctuations in the walls, but not in the domains. First, we note that the experimentally observed drop of the intensities to ca 15-20% during the first ps in Fig. 4a) corresponds to a fluctuation tilt angle of ca. 64-68° in Fig. 5b (dashed lines). We assume that during the following 100 ps the fluctuations in the walls reduce faster than in the domains caused by the above discussed temperature (and by that delay time) scaling of the order parameters such that finally the solid lines in Fig. 5b describe the intensities. At 100 ps the value of CL+CR observed in Fig. 4a) is 30% and CL-CR is 41% in good agreement with the numerical intensities at a tilt angle of 60° (solid lines in Fig. 5b).

Fig. 5c displays the evolution of the $2^{nd}$ moment of the scattering peak S(Q) from the simulation as a function of wall width. In contrast to the overall intensity (Fig. 5a) we do not observe significant changes in the second moments on increasing the relevant domain widths between 8 nm and 14 nm.

Fluctuations, however, can change the $2^{nd}$ moments significantly (Fig. 5d). While the $2^{nd}$ moments of the CL-CR exhibit a non-monotonous behaviour with tilt angle (red and orange lines), the CL+CR signal increases with increasing fluctuations - in good agreement with the experimental findings (Fig. 4d). Therefore, the experimentally observed decrease of the $2^{nd}$ moments of CL+CR after the first picoseconds can be explained by a reduction of the fluctuation strength with increasing delay time. This also holds true when including a faster reduction of fluctuations in the domain wall leading to a faster build-up of the chiral order.

Thus, the findings obtained from the second moments support the presence of the fluctuation mechanism (2). An additional contribution of the breathing mode mechanism (1) to the faster recovery of the total intensity cannot be be ruled out completely, since both mechanisms can in principle explain the observed faster recovery of the total intensity of CL-CR, although mechanism (1) overestimates the corresponding increase significantly. However, the presence of mechanism (1) implicates the necessity of breathing frequencies lower than 5 GHz, since otherwise also a decrease of the DW width and accordingly CL-CR should be observed within the 100 ps timeframe, which is not the case. Analytical calculations of the domain wall oscillation frequency [44] using the measured parameters suggest that even the strongly reduced anisotropy values at the high spin system temperatures after the IR excitation lead to oscillation frequencies in the range of 13-17 GHz. This does not fit the experimental data, that could only be explained by an oscillation of < 5 GHz. However, if the breathing mode is strongly damped, it is possible that the domain wall expands without breathing oscillations from the room temperature width of 8 nm to 11 nm. For an expansion from 8 nm to 11 nm we obtain an increase of the CL-CR/CL+CR ratio of 1.4, similar to the experimental increase of 1.3.

Therefore, we conclude that the main driver behind our experimental findings is likely mechanism (2) leading to a faster recovery of the chiral order in the domain walls in comparison to the ferromagnetic order in the domains. The responsible mechanism(s) might in addition be also material system dependent. So for the future, one can explore further multilayer stacks and the time dependence of the total intensity on even longer timescales to probe if one can identify additionally the effects of a damped breathing mode of the domain wall excited by the IR laser.

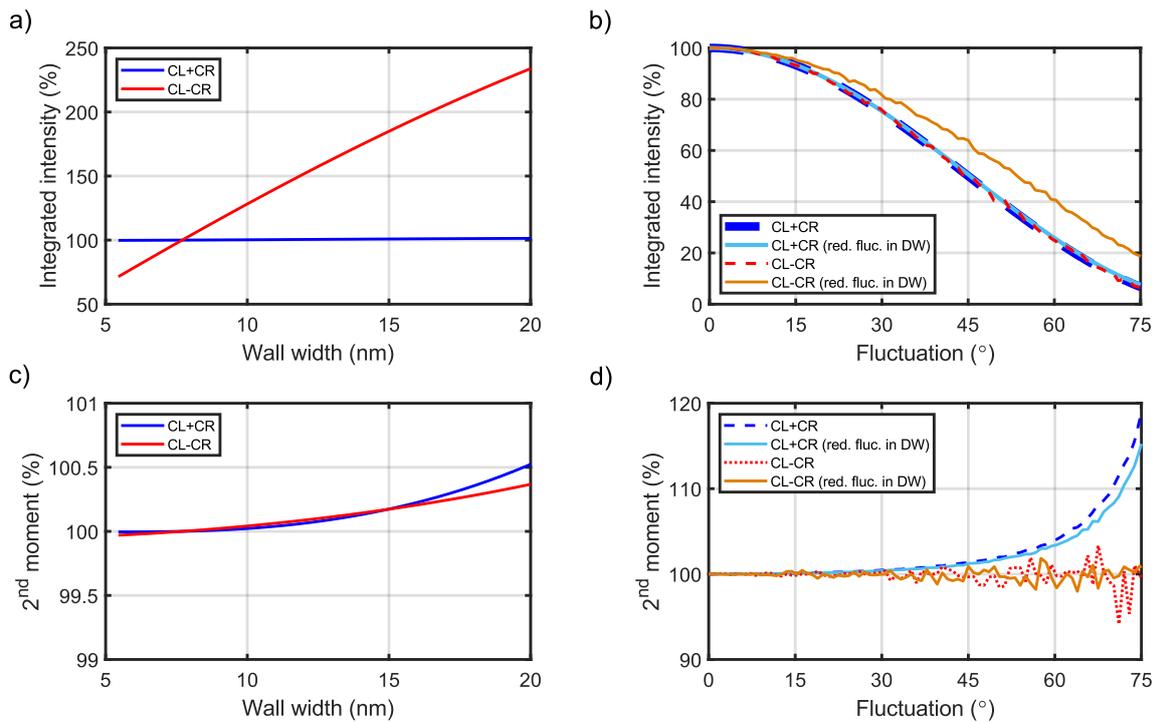

**Figure 5: Numerical calculations of total scattering intensities and second moments.** The numerically calculated (**a,b**) integrated intensity and (**c,d**) $2^{nd}$ moments of the sum signal (ferromagnetic order) and the difference signal (chiral magnetic order) as a function of the (**a,c**) domain wall width and (**b,d**) the transversal magnetisation fluctuation strength (cone angle in degrees). The 100 % value in (**a,c**) corresponds to the room temperature domain wall width of 8 nm. The dashed lines in (**b,d**) correspond to the case of fluctuation in the whole system, while the solid lines correspond to the case of reduced fluctuations within the domain walls.

In conclusion, using resonant magnetic scattering at an XUV FEL we identify a faster recovery of the scattering intensity signal stemming from the chiral magnetic order in the domain walls compared to the signal from the collinear magnetic order in the domains. We also observe an increase of the width of the structure factor peak for the collinear signal while the chiral scattering peak does not increase in width upon IR pumping indicating strong fluctuations. We explain both observations with the onset of strong transversal magnetic fluctuations after pumping in the domain systems. Importantly, however, our experimental findings imply reduced fluctuations within the chiral domain walls leading to a chiral order that is restored faster. We can connect this to the realisation that the characteristic time scale for the onset of the chiral magnetic order is significantly lower than the onset of the ferromagnetic order. In the future further fundamental aspects can be studied in detail e.g. the dependence of the timescales of the chiral order build-up on the absolute strength of the DMI by varying the heavy metal layers. The better control of the DMI and the chirality of spin structures on the ultrafast timescale can finally allow the controlled ultrafast manipulation of chiral magnetism, e.g. ultrafast writing of chiral topological objects such as skyrmions and pave the path to applications in the field of ultrafast chiral spintronics.

## Methods

**Sample fabrication and characterisation**

The magnetic multilayer sample has been grown at room temperature by dc magnetron sputtering and consist of (thicknesses in nm) Si/SiO$_2$/[Ta(5.3)/Co$_{20}$Fe$_{60}$B$_{20}$(0.93)/Ta(0.08)/MgO(2.0)]$_{x20}$/Ta(1.6). The Ta(0.08 nm) interlayer was inserted to tune the anisotropy stemming from the Co$_{20}$Fe$_{60}$B$_{20}$/MgO interface, while the 1.6 nm Ta capping was reduced to a thickness that prevented the sample from oxidation, but simultaneously led to sufficient penetration of the XUV light into the topmost Co$_{20}$Fe$_{60}$B$_{20}$ layer to ensure sufficient statistics. The magnetic properties of the sample were characterized using superconducting quantum interference magnetometry (SQUID) in the temperature range 10 K to 390 K and a vibrating sample magnetometer (VSM) in the temperature range 300 K to 600 K. Both out-of-plane and in-plane hysteresis loops were measured (see Supplementary S1) for each temperature. From those the saturation magnetisation $M_s$ and the effective perpendicular anisotropy constant $K_{eff}$ were determined. The latter was ascertained from the difference in the areas of the in-plane and out-of-plane loops [45]. This leads to the temperature dependencies shown in Supplementary S1 yielding a Curie temperature of (553±1) K and parameters $M_s$(300 K) = (844±28) kA/m and $K_{eff}$(300 K) = (133±17) kJ/m$^3$ at room temperature. Additionally, a 20x20 µm$^2$ MFM image of the multilayers was performed at room temperature as shown in Fig. 1b. The sample exhibits labyrinth-like stripe domains with a domain periodicity of (455+/-30) nm, as extracted from the Fourier spectrum which exhibits a first order peak at q = (13.8+/-0.9) µm$^{-1}$ (see Supplementary S5). The interfacial DMI arises from the Ta(5.3 nm)/Co$_{20}$Fe$_{60}$B$_{20}$ interface, which favours right-handed chiral Néel domain walls, as observed in the dichroic SAXS signal in Fig. 1e,f. One must take into consideration that the Ta(0.08 nm) insertion layer is not even a complete Ta monolayer and therefore a net interfacial DMI is still prominent in the sample. Finally, the interplay between interfacial DMI and dipolar interactions determines the equilibrium domain and domain wall arrangement in the sample [37] as revealed in the micromagnetic simulations of the full materials stack shown in Supplementary S3.

**Pump-probe experiments**

The measurements have been performed at the DiProI beamline [46-47] at the FERMI FEL facility in Trieste, Italy [48-49]. For the experiment, the FEL was tuned to the Fe M$_{2,3}$-edge at a wavelength of 23.0 nm, a pulse duration of 60 fs, a repetition rate of 50 Hz, and an attenuated pulse energy of 0.94 µJ. Using a Kirkpatrick-Baez (KB) optics, the beam was focused down to a size of (200) × (200) µm$^2$ leading to an energy density of 3,0 mJ/cm$^2$. The optical laser for pump-probe experiments is the same as the Ti:sapphire seed laser used for generating the FEL pulses in the HGHG scheme and therefore is intrinsically synchronized to the XUV-FEL pulses with a jitter of less than 10 fs. We used as a pump a 780 nm IR pulse of 100 fs duration, with an energy of 2.62 µJ and a size of (300) × (300) µm$^2$, leading to a pump energy density of 3.7 mJ/cm$^2$, well below the material damage threshold and the intra-pulse self- demagnetization process observed with FEL radiation on similar magnetic structures [50]. The IR and FEL beams are nearly parallel with a small angular offset of 2° which allows for a constant temporal resolution during the rotation of the sample through the beam. The resonant magnetic XUV reflectivity experiments were performed for left- and right-hand circular polarization with a charge-coupled device (CCD) area detector (2,048×2,048 pixels, 13.5×13.5 µm$^2$ pixel size) placed 145 mm behind the sample. The circularly polarized incident X-rays hit the sample at an angle of 44 degrees yielding an effective penetration depth of 8 nm. The intense reflected primary beam is blocked by a beam stop. The CCD is IR protected by an Al filter. Additional data obtained during the second experimental run and shown in Fig. 4b were measured at three different positions on the sample, under similar experimental conditions. The energy densities were about 3.8 mJ/cm$^2$ for pump (IR laser) beam and 2.0 mJ/cm$^2$ for probe (FEL) beam.

**SAXS data treatment**

For each measured pump probe time delay, 7000 SAXS patterns have been measured for each helicity. The patterns have been normalized to the incoming flux and averaged. The area around the beamstop and charge scattering streaks along the reflectivity ridge have been masked (white areas in Fig. 2b-2g) leaving $7.12 \times 10^5$ pixel for the magnetic scattering analysis. To remove the contribution of the charge component, the background signal is calculated for each pattern. Using the areas inside and outside the diffraction ring, we can reconstruct the whole pattern without the magnetic part. As a result of the background removal, we obtain the corrected patterns for post-proceeding analysis.

Afterwards the average intensity of the sum (CL+CR) and difference (CL-CR) diffraction images was determined as a function of the wavevector $Q_r$ for different time delays as depicted in Fig. 3. The radial profiles for each time delay were then integrated in order to obtain the evolution of the total intensity of the sum and the difference signal as a function of delay time. The time evolution of the total intensity normalized to the unpumped total intensity is displayed in Fig. 4 a & 4b.

The time dependence of the total intensities was fitted with the sum of three exponential functions convolved with a Gaussian distribution containing the time resolution $\sigma_t$ of the experiment given by the pulse duration of the FEL:

$$I(t)/I_{unpumped} = \left\{1 - H(t-t_0) \cdot \left[A_d \cdot \exp\left(-\frac{t-t_0}{\tau_d}\right) + A_{r1} \cdot \exp\left(-\frac{t-t_0}{\tau_{r1}}\right) + A_{r2} \cdot \exp\left(-\frac{t-t_0}{\tau_{r2}}\right)\right]\right\} * \left[\frac{1}{\sqrt{2\pi}\sigma_t} \cdot \exp\left(-\frac{(t-t_0)^2}{2\sigma_t^2}\right)\right], \quad (3)$$

where $\tau_d$ denotes the time constant of the demagnetisation respectively the dechiralisation process and $\tau_{r1}$ and $\tau_{r2}$ the time constants of the fast and slow recovery processes, whereas $A_d$, $A_{r1}$ and $A_{r2}$ denote the strength of these processes (fit parameters displayed in Supplementary S6). H(t) is the Heaviside step function and time zero $t_0$ was obtained from the fit.

The 2nd moments of scattering peak S(Q) (radial scattering distribution) displayed in Fig. 4c & 4d are calculated by:

$$2^{nd}\ moment = \frac{\int (q-q_0)^2\ S(q)\ dq}{\int S(q)\ dq} \quad (4)$$

where the peak center $q_0$ is determined for each delay time individually.

**Numerical calculation of the diffracted intensity**

In the simulation setup we assume a system of homochiral walls, where the components of the magnetization profiles of the -+ and +- walls are expressed by:

$$m_x^{W-+}(x) = -m_x^{W+-}(x) = -\sqrt{1 - [\tanh\frac{x}{w}]^2}$$
$$m_y^{W-+}(x) = m_y^{W+-}(x) = 0$$
$$m_z^{W-+}(x) = -m_z^{W+-}(x) = \tanh\frac{x}{w}, \quad (5)$$

with w being the domain wall width.

In a system of perfectly arranged parallel homochiral DWs with DW periodicity *d* the magnetization in the magnetic unit cell (from -*d*/2 to *d*/2) is given by

$$\boldsymbol{m}(x) = \begin{cases} \boldsymbol{m}^{W+-}\left(x+\frac{d}{4}\right) & \text{for } -\frac{d}{2} < x < 0 \\ \boldsymbol{m}^{W-+}\left(x-\frac{d}{4}\right) & \text{for } +\frac{d}{2} > x > 0. \end{cases} \quad (6)$$

The scattering amplitude resulting from n unit cells is given by

$$F(Q) = \sum_n e^{iQdn}(\hat{\varepsilon} \times \hat{\varepsilon}') \cdot \int_{-d/2}^{d/2} \boldsymbol{m}(x)e^{iQx}dx \quad ,\qquad(7)$$

where $\hat{\varepsilon}$ and $\hat{\varepsilon}'$ are the polarization vectors of the incident and scattered beam, respectively.
The form factor of the magnetic domains if given by the integral

$$\boldsymbol{m}(Q) = \int_{-\frac{d}{2}}^{\frac{d}{2}} \boldsymbol{m}(x)e^{iQx}dx =$$

$$= 2i\sin(Qd/4) \int_{-d/4}^{d/4} \boldsymbol{m}^{W-+}(x)\, e^{iQx}dx \qquad(8)$$

and is evaluated numerically. For circular polarized light of handedness λ the scattering intensity is

$$I(Q,\lambda) = \left|\sum_n e^{iQdn}\right|^2 |m_x(Q) + \lambda i m_z(Q)|^2 =$$

$$= \left|\sum_n e^{iQdn}\right|^2 \{|m_x(Q)|^2 + |m_z(Q)|^2 + 2\lambda[\Im m_x(Q)\Re m_z(Q) - \Re m_x(Q)\Im m_z(Q)]\} \qquad(9)$$

The difference between the CL and CR intensities is given by

$$I^{CL-CR}(Q) = 4\left|\sum_n e^{iQdn}\right|^2 [\Im m_x(Q)\Re f_z(Q) - \Re m_x(Q)\Im m_z(Q)] =$$

$$= 4\left|\sum_n e^{iQdn}\right|^2 \Im m_x(Q)\Re m_z(Q) \qquad(10)$$

while the sum of the CL and CR intensities is given by

$$I^{CL+CR}(Q) = 2\left|\sum_n e^{iQdn}\right|^2 [|m_x(Q)|^2 + |m_z(Q)|^2] \;. \qquad(11)$$

When including fluctuations we model the diffracted intensity that leads to the data shown in Fig. 5b and Fig. 5d numerically by calculating numerically the integrated intensity:

$$I(Q) = |(\varepsilon \times \varepsilon') \cdot \int_{-L/2}^{L/2} \boldsymbol{m}(x)e^{iQx}dx\,|^2, \qquad(12)$$

where L is the size of the simulation box. We used L=200 μm, corresponding to 444 domain periodicities of 450 nm. The magnetization **m**(x) in the system without fluctuations is obtained by alternating -+ and +- walls spaced 225 nm apart. The data shown in Fig. 5 is obtained by averaging over 100 random domain patterns.

In order to model the effect of transverse fluctuations we first tilt the magnetization at every point by a polar tilt angle. After that we rotate the magnetization at every point by an azimuthal angle obtained from a random number generator. We characterize the fluctuation strength by the polar tilt angle.

In order to model the case where fluctuations are reduced in the walls (due to the proposed faster recovery of the chiral order parameter) we switch off the fluctuations in the region $-1.1w < x - x_i < 1.1w$ for every wall i.

**Acknowledgments**

N.K., B.S., K.L., and M.K. gratefully acknowledge financial support by the Graduate School of Excellence Materials Science in Mainz (MAINZ, GSC266) and support from the Max Planck Graduate Center (MPGC). M.K. and the groups in Mainz acknowledge funding by the Deutsche Forschungsgemeinschaft (DFG, German Research Foundation)—projects 290319996/TRR173, 403502522/SPP 2137 Skyrmionics and 290396061/TRR173 and the EU (ERC-2019-SyG 3D MAGiC #856538). We acknowledge financial support from the Horizon 2020 Framework Programme of the European Commission under FET-Open Grant No. 863155 (s-Nebula). C.G. acknowledges funding by the Deutsche Forschungsgemeinschaft (DFG)—projects GU 535/4-1 and KS 62/1-1. Y.M. acknowledges funding from the DARPA TEE program through Grant MIPR (No. HR0011831554) from DOI and support from Leibniz Collaborative Excellence project OptiSPIN — Optical Control of Nanoscale Spin Textures. We also gratefully acknowledge the Jülich Supercomputing Center and RWTH Aachen University for providing computational resources under project jiff40. Measurements were carried out at the DiProI beamline at the FERMI FEL facility in Trieste, Italy. We thank FERMI FEL facility for the allocation of the beamtime and the technical support offered during the measurements. We acknowledge Felix Büttner for valuable discussions.

**Author Contributions**

C.G. and M.K. proposed the study. C.G., M.K., and Y.M. supervised the respective members of the study. N.K. and J.C. fabricated samples. N.K., C.G., H.Z., F.C., E.P. and I.L.-Q. performed the FEL experiment. D.K. and C.G. analysed the beamtime data. N.K., B.S., K.L. and D.L. performed the sample characterization. K.L. performed micromagnetic simulations. F.F., Y.M., N.K., C. G., M. K. and D.K. developed the two discussed mechanism. F.F. performed numerical calculations of the scattering signal. All authors participated in the discussion and interpreted results. N.K. drafted the manuscript with the help of D.K., F.F. and assistance from C.G., M.K. and Y.M. All authors commented on the manuscript.

**Competing interests**

The authors declare no competing interests.

**Data Availability Statement**

The data that support the findings of this study are available from the corresponding author upon reasonable request.

**Code Availability Statement**

The code that support the findings of this study is available from the corresponding author upon reasonable request.

## References

[1] Rößler, U. K., Bogdanov, A. N. & Pfleiderer, C. Spontaneous skyrmion ground states in magnetic metals. *Nature* **442**, 797–801 (2006).
[2] Mühlbauer, S. et al. Skyrmion lattice in a chiral magnet. *Science* **323**, 915–919 (2009).
[3] Bode, M. et al. Chiral magnetic order at surfaces driven by inversion asymmetry. *Nature* **447**, 190–193 (2007).
[4] Heinze, S. et al. Spontaneous atomic-scale magnetic skyrmion lattice in two dimensions. *Nat. Phys.* **7**, 713–718 (2011).
[5] Jiang, W. J. et al. Skyrmions in magnetic multilayers. *Phys. Rev.* **704**, 1–49 (2017).
[6] Everschor-Sitte, K., Masell, J., Reeve, R. M. & Kläui, M. Perspective: Magnetic skyrmions—Overview of recent progress in an active research field. *J. Appl. Phys.* **124**, 240901 (2018).
[7] Brataas, A. Chiral domain walls move faster. *Nat. Nanotech.* **8**, 485–486 (2013).
[8] Emori, S., Bauer, U., Ahn, S.-M., Martinez, E. & Beach, G. S. D. Current-driven dynamics of chiral ferromagnetic domain walls. *Nat. Mater.* **12**, 611–616 (2013).
[9] Fert, A., Cros, V. & Sampaio, J. Skyrmions on the track. *Nat. Nanotech.* **8**, 152–156 (2013).
[10] Woo, S. et al. Observation of room-temperature magnetic skyrmions and their current-driven dynamics in ultrathin metallic ferromagnets. *Nat. Mater.* **15**, 501–506 (2016).
[11] Dzyaloshinskii, I. E. A thermodynamic theory of weak ferromagnetism of antiferromagnets. *J. Phys. Chem. Solids* **4**, 241–255 (1958).
[12] Moriya, T. New mechanism of anisotropic superexchange interaction. *Phys. Rev. Lett.* **4**, 228–230 (1960).
[13] Tetienne, J.-P. et al. The nature of domain walls in ultrathin ferromagnets revealed by scanning nanomagnetometry. *Nat. Commun.* **6**, 6733 (2015).
[14] Pollard, S. D. et al. Observation of stable Néel skyrmions in cobalt/palladium multilayers with Lorentz transmission electron microscopy. *Nat. Commun.* **8**, 14761 (2017).
[15] Chen, G. et al. Out-of-plane chiral domain wall spin-structures in ultrathin in-plane magnets. *Nat. Commun.* **8**, 15302 (2017).
[16] Dürr, H. A. et al. Chiral Magnetic Domain Structures in Ultrathin FePd Films. *Science* **284**, 2166 (1999).
[17] Zhang, S. L. et al. Direct experimental determination of spiral spin structures via the dichroism extinction effect in resonant elastic soft x-ray scattering. *Phys. Rev. B* **96**, 094401 (2017).
[18] Chauleau, J.-Y. et al. Chirality in Magnetic Multilayers Probed by the Symmetry and the Amplitude of Dichroism in X-Ray Resonant Magnetic Scattering. *Phys. Rev. Lett.* **120**, 037202 (2018).
[19] Kim, K.-W. et al. Asymmetric skyrmion Hall effect in systems with a hybrid Dzyaloshinskii-Moriya interaction. *Phys. Rev. B* **97**, 224427 (2018).
[20] Vaterlaus, A., Beutler, T. & Meier, F. Spin-lattice relaxation time of ferromagnetic gadolinium determined with time-resolved spin-polarized photoemission. *Phys. Rev. Lett.* **67**, 3314–3317 (1991).
[21] Beaurepaire, E., Merle, J.-C., Daunois, A. & Bigot, J.-Y. Ultrafast spin dynamics in ferromagnetic nickel. *Phys. Rev. Lett.* **76**, 4250–4253 (1996).
[22] Rhie, H. S., Dürr, H. A. & Eberhardt, W. Femtosecond electron and spin dynamics in Ni/W(110) films. *Phys. Rev. Lett.* **90**, 247201 (2003).
[23] Stamm, C. et al. Femtosecond modification of electron localization and transfer of angular momentum in nickel. *Nat. Mater.* **6**, 740–743 (2007).
[24] Malinowski, G. et al. Control of speed and efficiency of ultrafast demagnetization by direct transfer of spin angular momentum. *Nat. Phys.* **4**, 855–858 (2008).
[25] Koopmans, B. et al. Explaining the paradoxical diversity of ultrafast laser-induced demagnetization. *Nat. Mater.* **9**, 259–265 (2010).
[26] Battiato, M., Carva, K., & Oppeneer, P. M. Superdiffusive Spin Transport as a Mechanism of Ultrafast Demagnetization. *Phys. Rev. Lett.* **105**, 027203 (2010).

[27] Pfau, B. et al. Ultrafast optical demagnetization manipulates nanoscale spin structure in domain walls. *Nat. Commun.* **3**, 1100 (2012).

[28] Sant, T. et al. Measurements of ultrafast spin-profiles and spin-diffusion properties in the domain wall area at a metal/ferromagnetic film interface. *Sci. Rep.* **7**, 15064 (2017).

[29] Rettig, L. et al. Itinerant and Localized Magnetization Dynamics in Antiferromagnetic Ho. *Phys. Rev. Lett.* **116**, 257202 (2016).

[30] Thielemann-Kühn, N. et al. Ultrafast and Energy-Efficient Quenching of Spin Order: Antiferromagnetism Beats Ferromagnetism. *Phys. Rev. Lett.* **119**, 197202 (2017).

[31] Schütz, G. et al. Absorption of circularly polarized x rays in iron. *Phys. Rev. Lett.* **58**, 737 (1987).

[32] Hannon, J. P. Trammell, G. T., Blume, M. & Gibbs, D. X-ray resonance exchange scattering. *Phys. Rev. Lett.* **61**, 1245–1248 (1988).

[33] Hellwig, O., Denbeaux, G.P., Kortright, J.B., Fullerton, E.E. X-ray studies of aligned magnetic stripe domains in perpendicular multilayers. *Physica B* **336**, 136–144 (2003).

[34] Zázvorka, J. et al. Thermal skyrmion diffusion used in a reshuffler device. *Nat. Nanotech.* **14**, 658–661 (2019).

[35] Conte, R. L. et al. Role of B diffusion in the interfacial Dzyaloshinskii-Moriya interaction in Ta/Co20Fe60B20/MgO nanowires. *Phys. Rev. B* **91**, 014433 (2015).

[36] Casiraghi, A. et al. Individual skyrmion manipulation by local magnetic field gradients. *Commun. Phys.* **2**, 145 (2019).

[37] Legrand, W. et al. Hybrid chiral domain walls and skyrmions in magnetic multilayers. *Sci. Adv.* **4**, eaat0415 (2018).

[38] Gutt, C. et al. Resonant magnetic scattering with soft x-ray pulses from a free-electron laser operating at 1.59 nm. *Phys. Rev. B* **79**, 212406 (2009).

[39] Villain, J. Two-level systems in a spin-glass model. I. General formalism and two-dimensional model. *J. Phys. C* **10**, 4793 (1977).

[40] Cinti, F. et al. Two-Step Magnetic Ordering in Quasi-One-Dimensional Helimagnets: Possible Experimental Validation of Villain's Conjecture about a Chiral Spin Liquid State. *Phys. Rev. Lett.* **100**, 057203 (2008).

[41] Menzel, M. et al. Information Transfer by Vector Spin Chirality in Finite Magnetic Chains. *Phys. Rev. Lett.* **108**, 197204 (2012).

[42] Harada, I. One-dimensional Classical Planar Model with competing interactions. *J. Phys. Soc. Jpn.* **53**, 1643 (1984).

[43] Wen, X., Wilczek, F., Zee, A. Chiral spin states and superconductivity. *Phys. Rev. B.* **39**, 11413 (1989).

[44] Liu, Y. & Grütter, P. Theory of magnetoelastic dissipation due to domain wall width oscillation. *J. Appl. Phys.* **83**, 5922 (1998).

[45] Lemesh, I., Büttner, F., & Beach, G. S. D. Accurate model of the stripe domain phase of perpendicularly magnetized multilayers. *Phys. Rev. B* **95**, 174423 (2017).

[46] Capotondi, F. et al. Multipurpose end-station for coherent diffraction imaging and scattering at FERMI@ Elettra free-electron laser facility. *J. Synchr. Rad.* **22**, 544(2015).

[47] Capotondi, F. et al. Coherent imaging using seeded free-electron laser pulses with variable polarization: First results and research opportunities. *Rev Sci Instrum* **84**, 051301 (2013).

[48] Allaria, E. et al. Highly coherent and stable pulses from the FERMI seeded free-electron laser in the extreme ultraviolet. *Nat. Photonics.* **6**, 699–704 (2012).

[49] Allaria, E. et al. Two-stage seeded soft-X-ray free-electron laser. *Nat. Photonics* **7**, 913–918 (2013).

[50] Müller, L. et al. Breakdown of the X-Ray Resonant Magnetic Scattering Signal during Intense Pulses of Extreme Ultraviolet Free-Electron-Laser Radiation. *Phys. Rev. Lett.* **110**, 234801 (2013).

# Supplementary

**S1: Determination of material parameters and their temperature dependence by SQUID/VSM**

Hysteresis loops were recorded by SQUID and VSM in order to determine the saturation magnetisation $M_s$ as well as the effective anisotropy $K_{eff}$ of the sample. The curves displayed in Fig. S1 show the out-of-plane (OOP) and in-plane (IP) hysteresis loops at 300 K. The effective perpendicular anisotropy $K_{eff}$ was determined by the difference in the areas of the in-plane and out-of-plane hysteresis loops and corresponds to $K_u - \mu_0/2\, M_s^2$ with the uniaxial anisotropy $K_u$ [1].

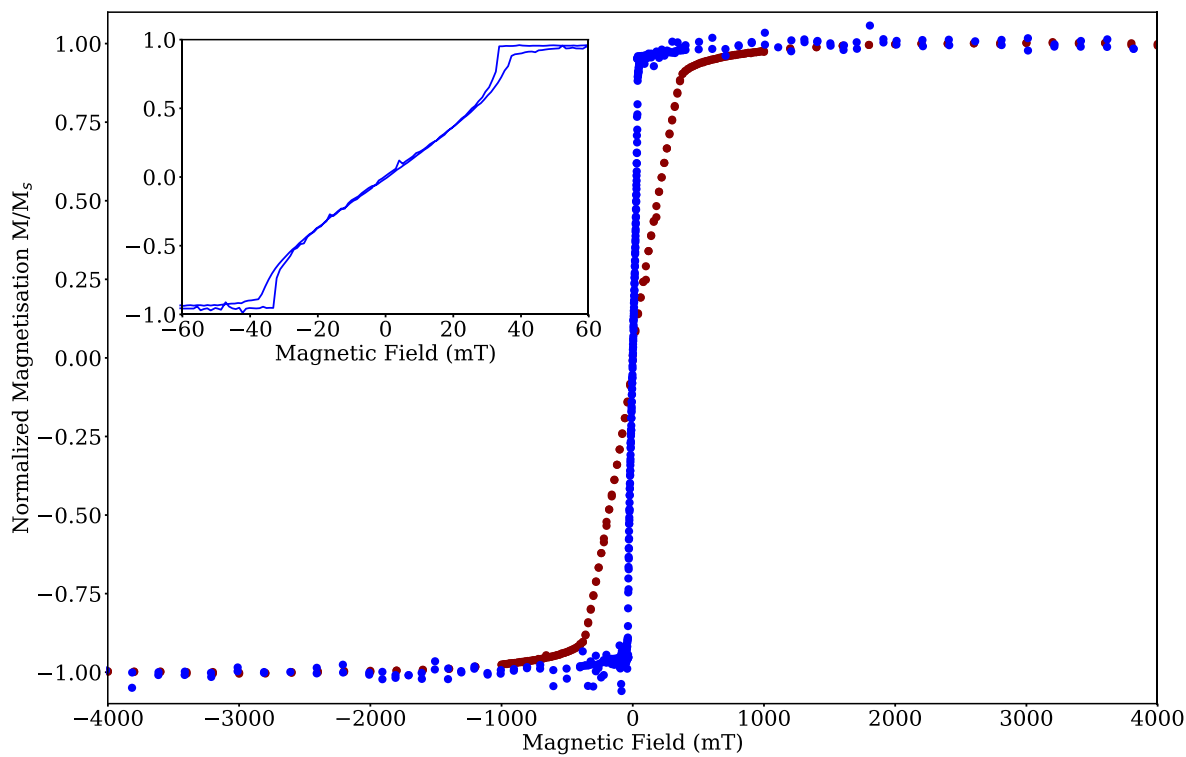

**Figure S1: Hysteresis loops of multilayer stack.** OOP (blue) and IP (red) hysteresis loops recorded at 300 K by SQUID magnetometry. Inset: The zoomed in OOP curve displays the typical behaviour of a multidomain stripe sample.

The hysteresis loops displayed in Fig. S1 measured at room temperature (300 K) via SQUID were also performed at temperatures of 10 K, 50 K, 100 K, 200 K, 300 K, 350 K and 390 K by SQUID and 300 K, 350 K, 390 K, 450 K, 500 K, 550 K, and 600 K by VSM. The saturation magnetisation was determined and afterwards the following model was used to fit the temperature dependence of the saturation magnetisation:

$$M_s(T) = M_s(0\ K)\,(1 - (T/T_c)^\alpha)^\beta, \quad (13)$$

with a saturation magnetisation $M_s(0\ K) = (976\pm10)$ kA/m, a Curie temperature $T_c = (553\pm1)$ K, $\alpha = (1.73\pm0.13)$ and $\beta = (0.63\pm0.04)$.

The temperature dependence of the effective anisotropy constant was fitted using:

$$K_{eff}(T) = K_u(T) - \frac{\mu_0}{2}(M_s(T))^2, \quad (14)$$

where the magnetisation dependence of the anisotropy constant was fitted by a power law:

$$\frac{K_u(T)}{K_u(0\ K)} = \left(\frac{M_s(T)}{M_s(0\ K)}\right)^c, \quad (15)$$

leading to an anisotropy constant $K_u$(0 K) = (1031±11) kJ/m³ and c = (2.16±0.02).

Using the fit parameters, we obtain finally a room temperature saturation magnetisation of $M_s$(300 K) = (844±28) kA/m and an effective anisotropy of $K_{eff}$(300 K) = (133±17) kJ/m³.

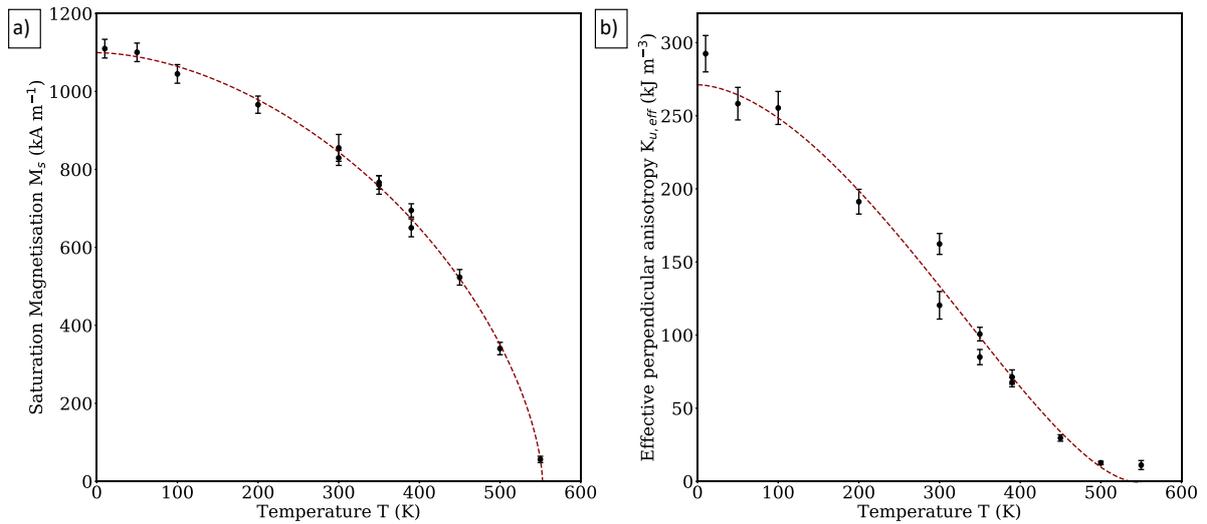

**Figure S2: Temperature dependence of $M_s$ and $K_{eff}$.** Temperature dependence of the saturation magnetisation $M_s$ (**a**) and effective anisotropy $K_{eff}$ (**b**). The datapoints were fitted by the models displayed in Eq. (13) respectively Eq. (14).

**S2: Pump-probe-delay time dependent spin system temperature**

The total intensity of the sum signal displayed in Fig. 4a & 4b is connected to average domain magnetisation (I ∝ $M^2$). Using the temperature dependence of the saturation magnetisation according to Fig. S2 we estimate the temperature evolution of the spin system during the pump-probe experiment displayed in Fig. S3. Upon optical excitation the spin system temperature increases within a ps to (493±22) K and relaxes afterwards slowly back to room temperature with an already decreased temperature of (469±13) K at 100 ps.

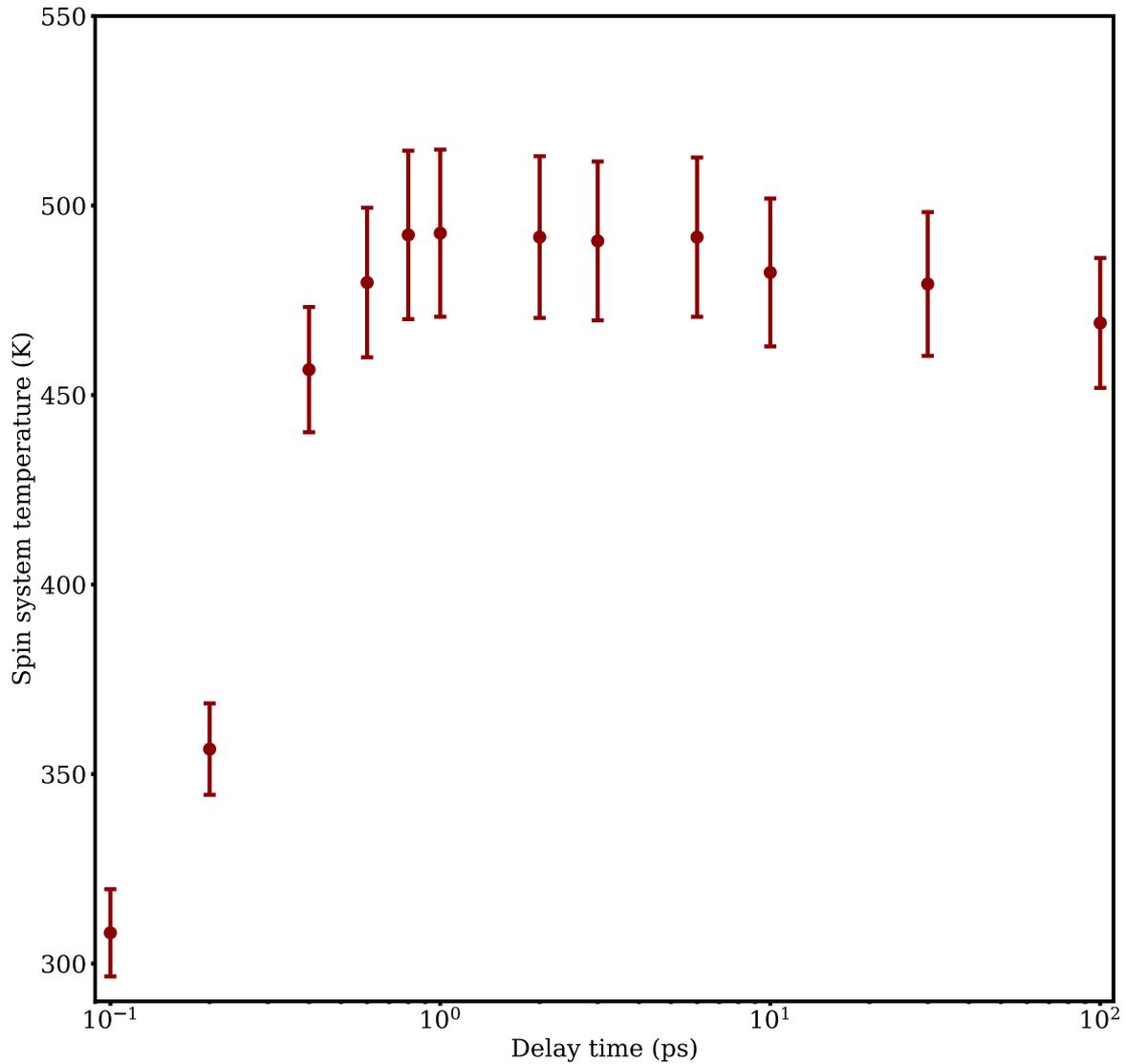

**Figure S3: Time dependence of spin system temperature.** The evolution of the spin system temperature as a function of pump-probe delay time in logarithmic scale.

## S3: Micromagnetic simulations of the equilibrium domain wall configuration

Using MicroMagnum [2], a framework to perform numerical micromagnetic simulations, we simulated the complete [Ta(5.3 nm)/Co$_{20}$Fe$_{60}$B$_{20}$(0.93 nm)/Ta(0.08 nm)/MgO(2.0 nm)]$_{x20}$ multilayer stack using a cell size of 0.9 nm. The following material parameters have been used for the magnetic layers: the experimentally measured parameters $M_s$(300 K) = 844 kA/m and $K_{eff}$(300 K) = 133 kJ/m$^3$, as well as the typical literature values of such material stacks for the DMI constant D = 0.06 mJ/m$^2$ [3-4] and the typically for CoFeB assumed exchange stiffness constant of A=10 pJ/m at room temperature [4]. The equilibration of a domain wall leads to the stabilization of the domain wall configuration as shown in Fig. S4. The exact domain wall arrangement is determined by an interplay between interfacial DMI and dipolar interactions that lead in the simulations to the stabilization of a so-called hybrid chiral domain wall [5]. The top and bottom magnetic layers in such a structure have opposite chirality due to flux closure, while the right-handed chirality is predominant in more layers due to the interfacial DMI provided by the Ta/CoFeB interface.

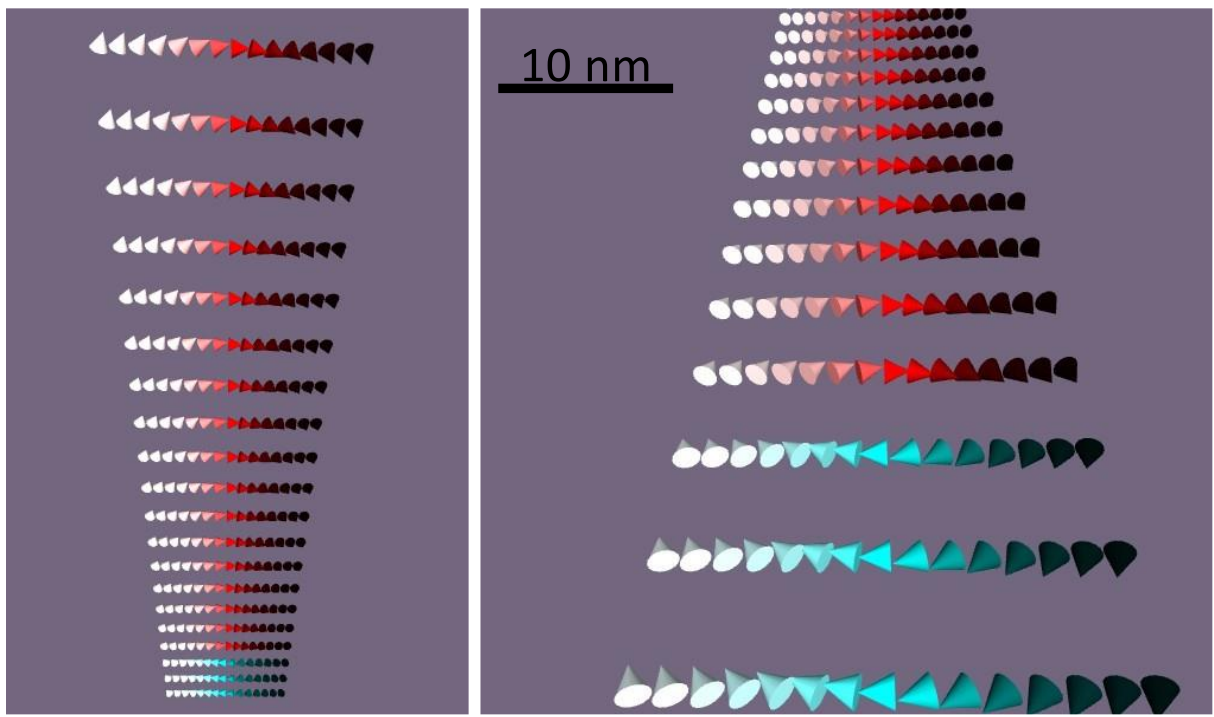

**Figure S4: Domain wall configuration of the complete material stack.** The magnetic moments of the individual cells of the different layers are depicted by the coloured arrows. While most layers are right-handed Néel type the three bottom layers are left-handed Néel type domain walls.

Analysing the domain wall profile in the simulations for the uppermost layer leads to a domain wall angle of φ = 90° degrees, supporting the experimental observation of fully right-handed Néel-type domain walls from the dichroic scattering signal.

By plotting the z-component of the magnetisation of the uppermost layer and using the expressions in Eq. (5) of the method sections, that are typically used to model homochiral domain walls, the domain wall width can be determined.

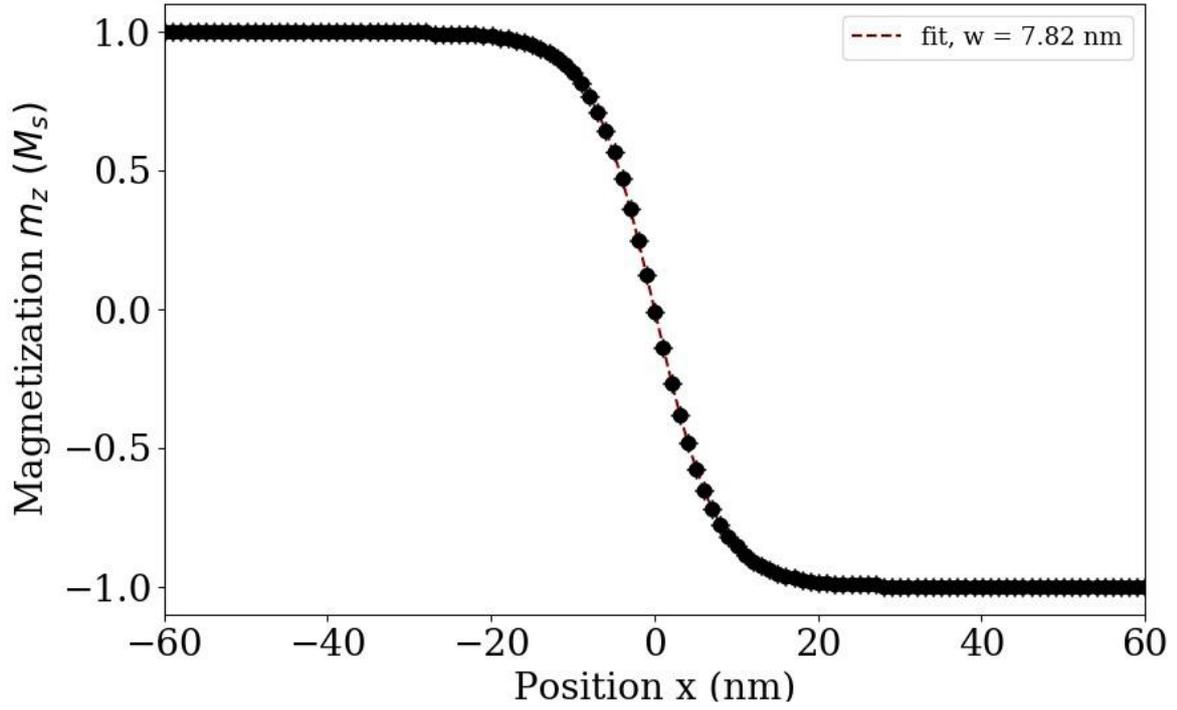

**Figure S5: Magnetisation profile of a domain wall in the topmost magnetic layer from micromagnetic simulation.** Fitting the profile of the z-component of the magnetisation leads to a domain wall width of approximately 8 nm.

Using the fit function $m_z^{W+-}(x) = -\tanh(\frac{x}{w})$, leads to a DW width of w = 7.82 nm ≈ 8 nm using the room temperature parameters.

This value is not differing significantly from the value obtained by the analytical formula:

$$\Delta = \sqrt{A/K_{u,eff}}, \quad (16)$$

typically used to determine the domain wall width from experimental material parameters [1]. Calculating the wall width with our material parameters leads to Δ = 8.6 nm. This shows that DMI as well as dipolar interactions contribute here strongly to the exact domain wall configuration [5], while the domain wall width is not affected significantly by the value of the DMI [4,6].

## S4: Temperature dependence of the domain wall width

The domain wall width in these thin film samples is governed by Eq. (16) with no significant effect of the DMI [4,6]. Using the temperature dependence of the effective anisotropy constant obtained in Fig. S2 one can estimate the temperature dependence of the (equilibrated) domain wall width using Eq. (16).

The temperature dependence of the exchange stiffness constant A is in mean-field approximation described by a power law $A \propto M_s^2$ leading to the temperature scaling of the domain wall width displayed in Fig. S6. One can observe that the domain wall width increases with increasing temperature due to a stronger scaling of the effective anisotropy constant in comparison to the exchange constant. It was also shown that fluctuation corrections from nonlinear spin-wave effects lead to a scaling $A \propto M_s^\kappa$ with κ<2 depending on the lattice structure [7], which would lead to an even stronger scaling of the domain wall width with the temperature.

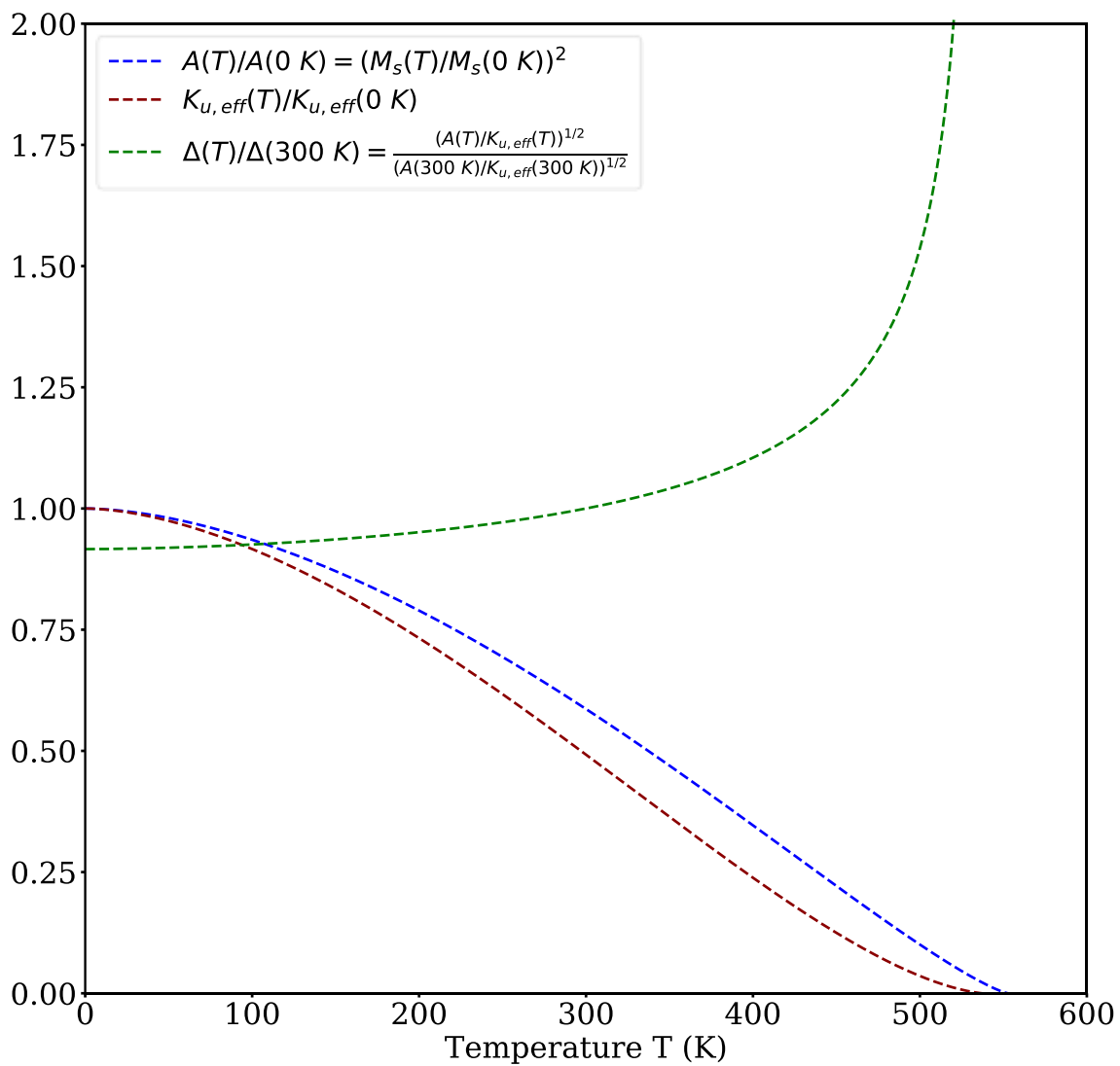

**Figure S6: Temperature dependence of $K_{eff}$, A and Δ.** Normalized temperature dependence of the effective anisotropy constant (red), the exchange stiffness constant (blue) and the analytical domain wall width Δ (green).

Using again the temperature scaling of the exchange stiffness constant and the effective anisotropy in combination with the domain wall width w(300 K)= 8nm obtained from the micromagnetic simulations and Eq. (16), leads to the domain wall widths displayed in Fig. S7.

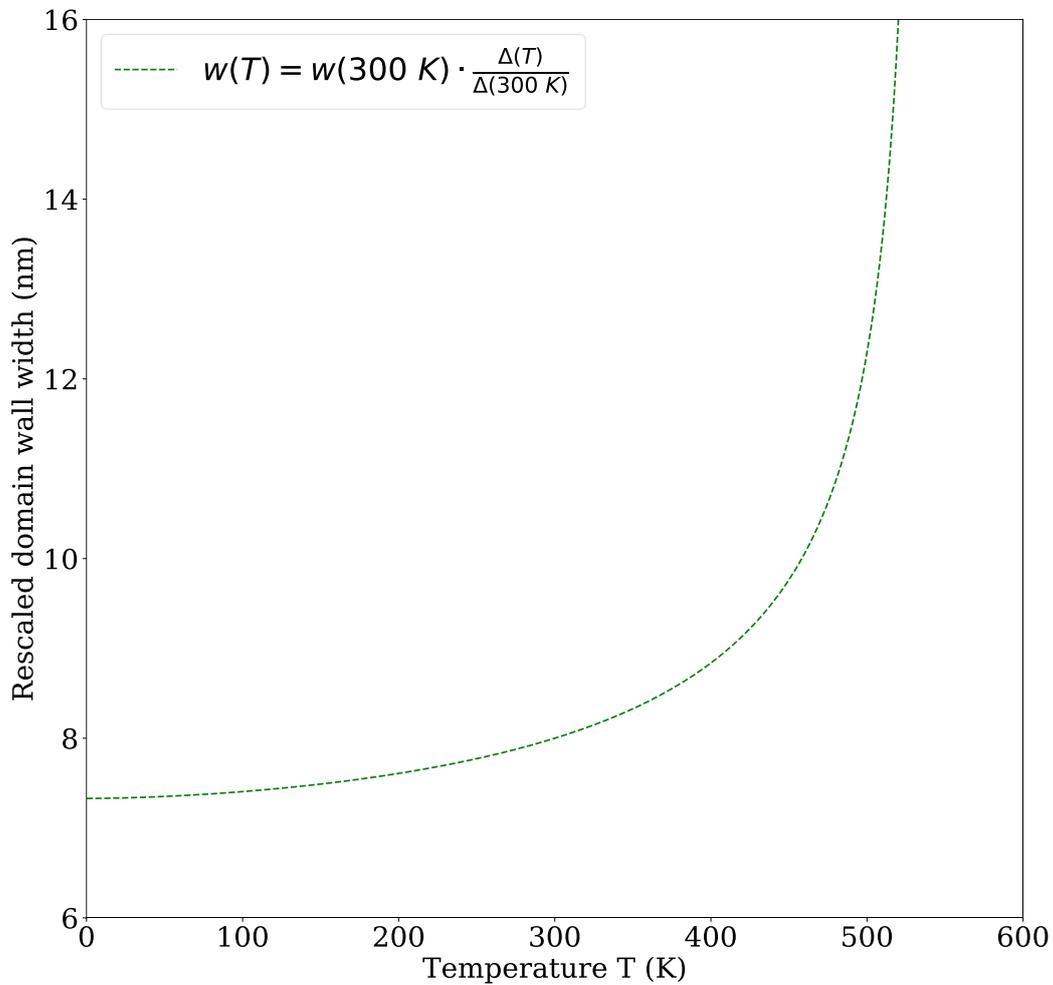

**Figure S7: Temperature dependence of w.** Temperature dependence of the domain wall width w based on the temperature scaling of the exchange stiffness constant and effective anisotropy.

Combining the time delay dependence of the spin system temperature displayed in Fig. S3 with the here obtained temperature dependence of the domain wall width one can estimate the equilibrium domain wall width at increased temperature for delay times between 1 ps to 100 ps to approximately 11 nm.

**S5: Fourier spectrum of the MFM image**

The magnetic force microscopy image displayed in Fig. 1b was Fourier transformed leading to the spectrum displayed in Fig. S8. One can see a clear first order peak at a value of $q=(13.8+/-0.9)$ µm$^{-1}$ as well as a second order peak at $q=(27.6+/-0.9)$ µm$^{-1}$ and a third order peak at $q=(41.4+/-1.0)$ µm$^{-1}$ due to partial short range stripe alignment of the OOP demagnetized magnetic labyrinth pattern. The first order peak indicates a stripe periodicity of (455+/-30) nm, which agrees well with the real space stripe data measured.

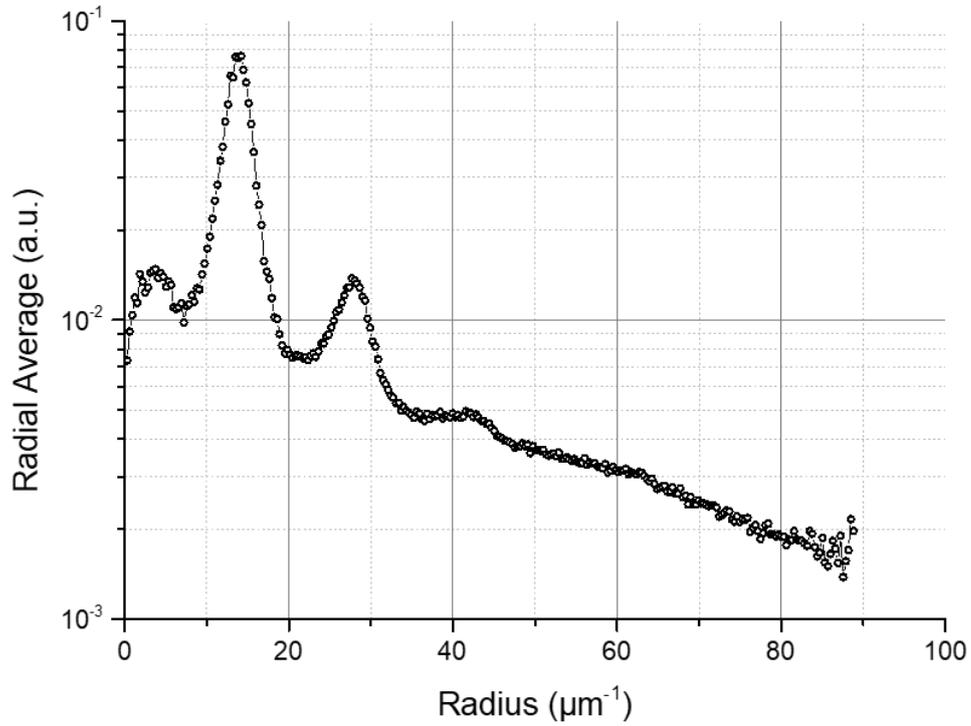

**Figure S8: Fourier Spectrum of MFM image.** Logarithmic Fourier spectrum of the MFM image of the demagnetized labyrinth stripe state showing a first, second and third order peak.

## S6: Fit parameters of the pump-probe delay curves from 0 - 3 ps

|  | Figure 4a | | Figure 4b | |
|---|---|---|---|---|
|  | *CL+CR* | *CL-CR* | *CL+CR* | *CL-CR* |
| $A_D$ | -0.80±0.01 | -0.85±0.03 | -0.86±0.01 | -0.89±0.01 |
| $\tau_D$, ps | 0.39±0.10 | 0.31±0.10 | 0.16±0.01 | 0.16±0.01 |
|  |  |  |  |  |
| $A_{R1}$* | *~0.04* | *~0.10* | 0.05±0.01 | 0.05±0.02 |
| $A_{R2}$* | *~0.76* | *~0.80* | 0.78±0.01 | 0.81±0.02 |
| $\tau_{R1}$ *, ps | *~1±1* | *~1±1* | *14±7* | *9±6* |
| $\tau_{R2}$, ps | >900 | 312±18 | >2000 | 530±100 |

*) Some of the refinement parameters are difficult to determine from the fit due to an ill-conditioned Jacobian.

**Table S1: Pump-probe fit parameters.** Fit parameters obtained by fitting the pump-probe data shown in Fig. 4a & 4b with the model described in Eq. (3).

## S7: Ultrafast demagnetisation within the first picoseconds

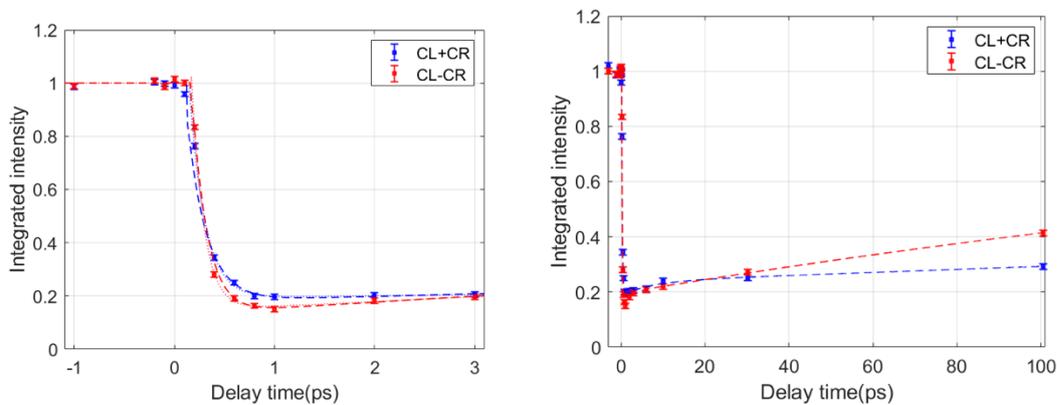

**Figure S9: Fits of the pump-probe delay curves.** Left: Ultrafast demagnetization (same data as in Fig. 4(a)) - solid lines represents a fit with Eq. (3) in the time interval t = (0-3) ps. Right: Solid line represents a fit over the entire time interval with Eq. (3).

## S8: Estimation of error bars

There are the following sources of fluctuations contributing to the error bar:

- Poisson noise from photon counting statistics. The CCD scattering images (see Fig. 1 and 2) represent averages of 7000 FEL pulses each - with the CCD camera read out after the 7000 pulses. The data points shown in Fig. 4a and 4b have been determined by azimuthal and radial averaging over 7.12x10$^5$ pixels in the relevant Q-range. The counting error based on Poisson statistics is $\Delta I = \sqrt{I}$ yielding a relative error of $\Delta I/I = 1/\sqrt{I}$. The averaged count numbers are rather high in the range of 0.1 to 1x10$^6$ photons implying relatively small error bars. We find typical values of $\Delta I/I$= 0.3 % at t = 1ps and $\Delta I/I$= 0.2 % at t=100 ps for CL-CR and 0.16 % (1ps) and 0.13% (100 ps) for CL+CR.
- FEL fluctuations. The FERMI FEL is a seeded FEL and intensity fluctuations were relatively small with standard deviation of 10-20 % during the experiment. With 7000 shots the error is std/sqrt(7000) ranging around $\Delta I/I$=0.1- 0.2 %.
- Speckle fluctuations. The use of a highly coherent beam results in a speckle pattern, visible by the grainy appearance of the CCD images and also by the non-flat appearance of the scattering curves in Fig. 3. The fact that they appear always at the same location in reciprocal space demonstrates the speckle origin of these features. The graininess of a speckle pattern can also yield fluctuations in the average intensity with values depending on the number of speckles N and external fluctuations produced by the pointing instability of the FEL (around 1/3 to 1 sigma of beam, FWHM). The reader may think about the extreme case if a single speckle is fluctuating. We estimate the magnitude of this contribution in the following way: The speckle size is given by $S = L\lambda/D$ with L denoting the sample-detector distance, $\lambda$ the wavelength and D the beam size. The number of speckles in the Q-area of interest between Q$_2$ and Q$_1$ is $N = \frac{\pi(L\lambda)^2(Q_2^2 - Q_1^2)}{(4\pi)^2 S^2}$. From the central part of I(Q) with Q=14.00 to 15.00 μm$^{-1}$ we obtain N=1000 and thus estimate an upper bound for a statistical fluctuation of $\Delta N/N \sim 0.65$ %.

Combining all sources of fluctuations result in the error bars shown in the manuscript with values of around 1%. We can compare this number to the experimental data by calculating the standard deviation of the first five measured data points in the time interval t=-3,-1,-0.2,-0.1,0 ps before pumping (=constant intensity). The standard deviation for CL+CR and CL-CR are 1.3 % each, which is in reasonable agreement with the values calculated above.

## S9 Additional Scans

In a second round of experiments we measured with lower scattering intensity three pump-probe scans. The data shown in Fig. 4b represent the average of these three scans.

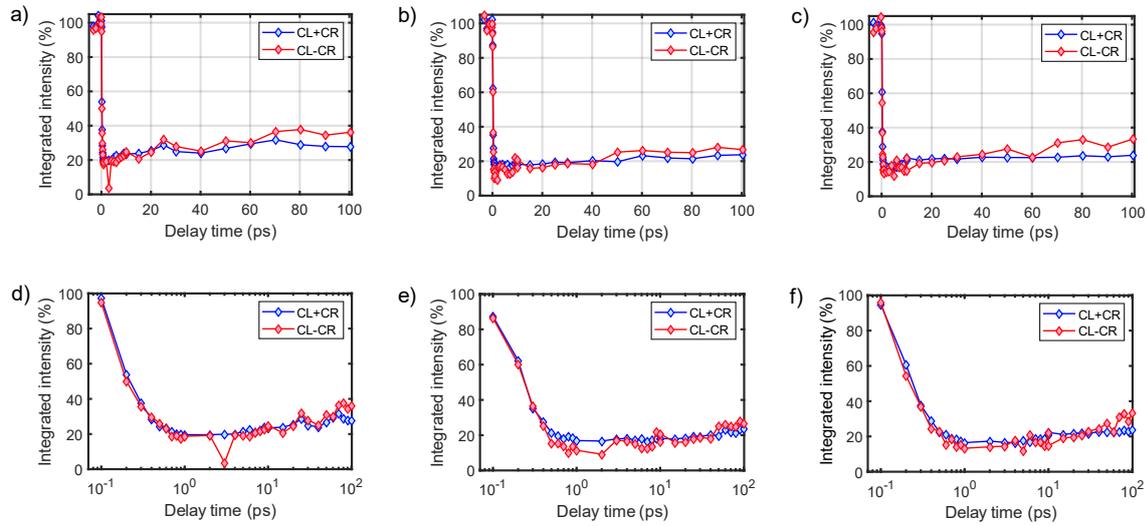

**Figure S10: Additional scans including more data points at different times.** Three additional scans with lower scattering intensity and more time points.

## References

[1] Lemesh, I., Büttner, F., & Beach, G. S. D. Accurate model of the stripe domain phase of perpendicularly magnetized multilayers. *Phys. Rev. B* **95**, 174423 (2017).
[2] https://github.com/MicroMagnum/MicroMagnum
[3] Conte, R. L. et al. Role of B diffusion in the interfacial Dzyaloshinskii-Moriya interaction in Ta/Co20Fe60B20/MgO nanowires. *Phys. Rev. B* **91**, 014433 (2015).
[4] Casiraghi, A. et al. Individual skyrmion manipulation by local magnetic field gradients. *Commun. Phys.* **2**, 145 (2019).
[5] Legrand, W. et al. Hybrid chiral domain walls and skyrmions in magnetic multilayers. *Sci. Adv.* **4**, eaat0415 (2018).
[6] Thiaville, A. et al. Dynamics of Dzyaloshinskii domain walls in ultrathin magnetic films. *EPL* **100**, 57002 (2012).
[7] Atxitia, U. et al. Multiscale modeling of magnetic materials: Temperature dependence of the exchange stiffness. *Phys. Rev. B* **82**, 134440 (2010).